\documentclass[prb,twocolumn,showpacs,amsmath,amssymb]{revtex4}

\usepackage{graphicx}

\begin{document}



\title{Disappearance of the de Almeida-Thouless line in six dimensions}


\author{M. A. Moore and A. J. Bray}

\affiliation{School of Physics and Astronomy, University of Manchester, 
Manchester M13 9PL, UK}

\date{\today}

\begin{abstract}
We  show that  the Almeida-Thouless  line in   Ising
spin  glasses  vanishes  when  their  dimension  $d\rightarrow  6^+$  as
$h_{AT}^2/T_c^2  =  C(d-6)^4(1- T/T_c)^{d/2  -  1}$,  where  $C$ is  a
constant of  order unity.  It is shown that  replica  symmetry  
breaking   also  stops   as  $d\rightarrow 6^+$. Equivalent results that  
could be  checked by  simulations are given for the  one-dimensional 
Ising  spin glass  with long-range interactions.  

\end{abstract}

\pacs{75.50.Lk, 75.40.Cx, 05.50.+q}

\maketitle



\section{Introduction}

The field  of spin glasses  is now well  into its fourth  decade, with
many important  questions still unresolved.  Chief among these  is the
nature of  the ordered phase. The  principal rival theories  of it are
(i)   the    replica-symmetry-breaking   (RSB)   theory    of   Parisi
\cite{Parisi},    motivated   by   the    exact   solution    of   the
Sherrington-Kirkpatrick   (SK)   mean-field   model,  and   (ii)   the
droplet/scaling  theory  \cite{Fisher-Huse,Bray-Moore}  based  on  the
properties  of excitations  in the  ordered phase.  There is  still no
general consensus on which (if any) of these theories is correct.

An  important (and  perhaps  the simplest)  discriminator between  the
theories is the predicted behavior  of the system when the temperature
is decreased in  the presence of an applied  magnetic field. According
to the  RSB scenario,  there will  still be a  phase transition  in an
applied field $h$, ocurring at a temperature $T_c(h)$, the 
Almeida-Thouless (AT)  line \cite{AT},  which decreases as  the field
increases. This line can be explicitly calculated in the mean-field SK
model. For small  $h$ it has the form $T_c(h) =  T_c(0) - {\rm const.}
h^{2/3}$ or, equivalently,
\begin{equation}
h_{AT}^2 \propto [T_c - T]^3\ ,
\label{AT}
\end{equation} 
($T_c(0)\equiv T_c$).  In  the RSB  theory, this  line is  the phase
boundary  between the (high-temperature)  replica symmetric  phase and
the (low-temperature) broken replica symmetry spin glass phase. Within
the droplet/scaling theory, on the  other hand, there is no such line:
An applied magnetic field is  predicted to remove the phase transition
completely and the low-temperature phase in the absence of a field has
replica symmetry.

Before presenting our calculation, we  recall that the {\em shape} of
the AT line  starts to differ from Eq.\  (\ref{AT}) already for $d<8$,
as shown  by Green et  al.\ \cite{Green-Moore-Bray} and by  Fisher and
Sompolinsky  \cite{Fisher-Sompolinsky}.  Instead  of Eq.\  (\ref{AT}),
these authors show that the AT line has the form
\begin{equation}
h_{AT}^2 \propto [T_c-T]^{d/2 - 1}\
\label{AT1}
\end{equation}
for  $6<d<8$.  Note  that as  $d  \to  8^-$  in Eq.\,(\ref{AT1}), 
Eq.\,(\ref{AT}) is recovered.  

In this paper  we will derive an  exact result for the form  of the AT
line  in  $d=6+\epsilon$ dimensions,  correct  to leading  non-trivial
order  in $\epsilon$.  The  result can  be written,  for $T$  close to
$T_c(0)$, as
\begin{equation}
\frac{h_{AT}^2}{T_c^2} = C(d-6)^4\left(1- \frac{T}{T_c} \right)^{d/2 -
1}\ ,
\label{final result}
\end{equation}
where $C$ is  a non-universal constant.  The form is  the same as that
proposed  by Green  at al.\  and by  Fisher and  Sompolinsky,  but the
amplitude contains  the factor $(d-6)^4$ which  vanishes (rapidly) for
$d \to 6$.

In itself  this result does not prove  that there is no  AT line below
six  dimensions. If there  were an  AT line  below six  dimensions, it
would   have  the  scaling form  $h_{AT}^2  \sim  \left(1-
T/T_c \right)^{\beta+\gamma}$, where  $\beta$ and $\gamma$ are
the    critical    exponents   of    the    zero-field   spin    glass
\cite{Fisher-Sompolinsky}.   The  vanishing   of   the  amplitude   in
Eq.  (\ref{final result})  would  arise to  ensure  continuity in  six
dimensions of  the forms above and  below $T_c$ \cite{APY}.  
Hence  we need to provide an  additional argument why
there should be no AT line when $d \le 6$.

Even before  the droplet/scaling theory  had been developed,  Bray and
Roberts  \cite{Bray-Roberts}  (BR)  had used  standard  renormalization
group (RG)  methods to  investigate the putative  RG fixed  point that
controls the  critical behavior at the  AT line. In  zero field there
are three degenerate soft modes  at the critical point (usually called
the  ``longitudinal'',   ``anomalous''  and  ``replicon''   modes).  A
conventional  RG analysis  \cite{Harris-Lubensky-Chen} shows  that the
upper  critical dimension  is $d_u=6$.  For $d>6$  the  Gaussian fixed
point  is stable  and  the critical  exponents  take their  mean-field
values.  For   $d<6$,  a  non-trivial   fixed  point  is   stable  and
non-classical exponents are obtained and  can be calculated as a power
series in $(6-d)$ in the conventional way.

An  applied magnetic  field,  however, changes  everything. The  field
lifts  the  degeneracy,  leaving  a  single soft  mode,  the  replicon
mode. In  their RG calculation, BR  discarded the two  hard modes, and
derived RG recursion relations appropriate  to the soft modes which we
discuss below. Again, the critical dimension is $d_u=6$. There are now
two coupling constants  $w_1$ and $w_2$.  In dimensions  $d>6$, the RG
flows have a single stable  fixed point, the Gaussian fixed point with
$w_1^*=w_2^*=0$, implying  that for $d>6$ there is  a continuous phase
transition, with critical exponents  given by their mean-field values.

For $d\le 6$,  however, no physical stable points  could be found.  BR
suggested that  this might be because  there was no AT  line for $d\le
6$.  We shall strengthen this  argument by examining the 
RG flows when $d>6$ in  more detail.  We  find that the  basin of 
attraction  of the Gaussian fixed point  shrinks to zero as 
$\epsilon  \rightarrow 0$; it has a linear extent of  order $\epsilon^{1/2}$.  
It thus seems natural to expect that there is no AT line when $d\le 6$: 
There is no physical stable fixed point  when $d\le 6$ and the basin  
of attraction for the controlling  fixed point  in $6+\epsilon$  
dimensions is  shrinking to zero as $\epsilon \rightarrow 0$.

Above   the   AT   line   the  high-temperature   phase   is   replica
symmetric. Below  it, the phase  has broken replica symmetry.  Thus if
the  AT line  is  disappearing as  $d\rightarrow 6^+$,  then one  would
naturally expect that replica symmetry breaking in the zero-field case
would vanish as  $d\rightarrow 6^+$. We shall demonstrate  that is indeed
the case by showing that the \lq\lq breakpoint'' $x_1$ in Parisi's RSB
function $q(x)$  \cite{Parisi} goes  to zero in  this limit.  Thus the
low-temperature phase for $d \le 6$ should be replica symmetric.

The structure of the paper is as follows. In section II, we present an RG 
analysis, valid for $d \ge 6$, that leads directly to 
Eq.\ (\ref{final result}).  In section III we analyse the consequences 
of our RG results for the form of the breakpoint, $x_1$, in the Parisi 
RSB theory. We find the $x_1$ tends to zero for $d \to 6^+$, suggesting 
that replica symmetry breaking goes away in six dimensions. In section IV, 
the RG equations of BR are presented. The basin of attraction of the 
Gaussian fixed point is determined numerically. It is a compact region 
enclosing the origin, with linear dimensions of order $\sqrt{\epsilon}$, 
shrinking to a point as $d \to 6^+$. In section V, we treat the 
one-dimensional spin-glass with long-range inetractions falling off 
with distance $r$ as $r^{-\sigma}$. This model has mean-field behavior 
for $\sigma < 2/3$ and non-classical behavior for $\sigma > 2/3$. 
Thus $\sigma<2/3$ corresponds to $d>6$ in the short-ranged model. 
We show that the de Almeida-Thouless line goes away as $\sigma \to 2/3^-$, 
and that the breakpoint $x_1$ of the Parisi function goes to zero in this 
limit. We conclude with a brief discussion of the main points in 
section VI. 
 
\section{Renormalization Group Analysis}

We now describe the calculations  that lead to the result quoted above.. 
We start from the  Ginzburg-Landau-Wilson free-energy functional  for the Ising
spin  glass which,  written in  terms of  the replica  order parameter
field, is
\begin{eqnarray}
&F[\{Q_{\alpha\beta}\}]  =   \int   d^dx\,   \left[\frac{1}{2}r
\sum_{\alpha<\beta}Q_{\alpha\beta}^2
+\frac{1}{2}\sum_{\alpha<\beta}(\nabla
Q_{\alpha\beta})^2\right. \nonumber  \\
&\hspace*{-0.2cm}  +   \left.\frac{w}{6}
\sum_{\alpha<\beta<\gamma}Q_{\alpha\beta}
Q_{\beta\gamma}Q_{\gamma\alpha}  -  h^2
\sum_{\alpha<\beta} Q_{\alpha\beta} + O(Q^4)\right]
\label{FQ}
\end{eqnarray}
where  $h$  is the  applied  field. We use conventional  RG    methods
\cite{Harris-Lubensky-Chen}, but  work {\em above}  the upper critical
dimension, $d_u=6$, so  we define $\epsilon = d-6$.   A simple scaling
analysis of the  terms in the functional of  Eq. (\ref{FQ}) shows that
the  natural size  of $h^2$  is $\sim  |r|^2/w$; this  remains the
correct scaling form  for all $d>6$. $h_{AT}^2$ is given by
Eq. (\ref{final  result}) and Eq.  (\ref{srfinal}) below,  and is much
smaller than  $|r|^2/w$ provided $w^2|r|^{\epsilon/2}\ll  1$. We shall
 work in  this limit  as it  allows the  use of  the  simple RG
equations of the  $h=0$ theory to obtain the  AT line as $T\rightarrow
T_c$. When $d\le 6$ this will not be possible and
 the full set of RG 
equations in  Ref. \cite{Temesvari} would have to be solved instead.

The renormalization group flows for $w$ and $r$ read \cite{Temesvari},
\begin{eqnarray}
\frac{dw}{dl} & = & \frac{1}{2}\left[-\epsilon -3\eta(l)\right]w-2w^3
\label{w-eq} \\
\frac{dr}{dl} & = & [2-\eta(l)]r - 4w^2 r,
\label{r-eq}
\end{eqnarray}
while $h^2$ evolves according to
\begin{equation}
\frac{d(h^2)}{dl} = \frac{d+2-\eta(l)}{2}\,h^2\ ,
\label{h-eq}
\end{equation}
where $\eta(l) = - (2/3)w(l)^2$.
The  RG equations  are correct  to $O(w^3)$,  and to  linear  order in
$r$. There  is an  additional $r$-independent term  of order  $w^2$ in
Eq.\ (\ref{r-eq}) that we have  omitted since it ultimately just leads
to a shift in $r$, i.e.\ a shift in the critical temperature. In these equations  the   usual   geometrical   factor
$K_d=2/\Gamma(d/2)(4\pi)^{d/2}$  has been  absorbed  into $w^2$.   For
$d>6$ the  Gaussian fixed point  is stable and the  critical exponents
take their mean-field values. 

Integrating the RG flow equations when $d>6$ up to scale $l$ gives
\begin{eqnarray}
w(l)                    &=&                   \frac{w(0)\exp(-\epsilon
l/2)}{\left[(2w(0)^2/\epsilon)(1-e^{-\epsilon l})+1\right]^{1/2}}\,
 \label{wsquared(l)} \\
r(l) & = &  r(0)\,\exp[2l -(10/3)\Delta(l)]\ , \label{r(l)}\\ h(l)^2 &
= & h(0)^2\,\exp[\{(d+2)/2\}l + (1/3)\Delta(l)]\ ,
\label{hsquared(l)}
\end{eqnarray}
where
\begin{equation}
\Delta(l)  =  \int_0^l\,  w(l')^2\,{\mathrm  d}l'  =  \frac{1}{2}  \ln
\left[\frac{2w(0)^2}{\epsilon}\left(1-e^{-\epsilon l}\right)+1\right]\,
\label{Delta(l)}
\end{equation}
and $w(0)=w$, $h^2(0)=h^2$, $r(0)=r$.

At large $l$,
\begin{equation}
 w(l)\rightarrow
 \left[\frac{\epsilon}{2(1+\epsilon/2w(0)^2)}\right]^{1/2}\exp(-\epsilon
 l/2).
 \label{largel}
 \end{equation}
 The  exponential follows  from the  $O(w)$ terms  in the  RG equation
(\ref{w-eq}). The $O(w^3)$ terms in Eq.\ (\ref{w-eq}) serve to fix the
amplitude of the asymptotic decay.  The basic idea is to integrate the
RG equations to a scale  $l^*$ at which the running coupling constant,
$w(l^*)$, is small enough for one-loop order perturbation theory to be
accurate.

The one-loop perturbative calculation of  the AT line has been carried
out by Green et al..  The result is \cite{Green-Moore-Bray}
\begin{equation}
h^2/Q = 144 w^2 |r|^2 I_d\ ,
\label{jenny}
\end{equation}
where  $I_d$  is  the  integral  $\int_0^\infty  d^dq\,  q^{-4}(q^2  +
|r|)^{-2}$, which equals $A_d|r|^{d/2-4}$ and $A_d=1/2$ for $d=6$.  In
Eq.\  (\ref{jenny}),  the  geometrical  factor $K_d$  has  again  been
absorbed  into $w^2$, and  $Q$ is  the mean-field  value of  the order
parameter for $r<0$. The factor of  36 between this result and that in
\cite{Green-Moore-Bray} is  due to  the different definitions  of $w$,
which differ  by a  factor of  6 in the  two calculations.   Using the
result $Q \sim 3|r|/w$, which holds on the AT line 
 in 
the perturbative limit where $w^2|r|^{\epsilon/2}\ll 1$, we find
\begin{equation}
h^2 \sim w|r|^{d/2-1},\ \ \ (6 < d <8),
\label{GBMFS}
\end{equation} 
which  is the  result obtained  in Refs.\  \cite{Green-Moore-Bray} and
\cite{Fisher-Sompolinsky}.  

To exploit  the perturbation theory  result, we coarse-grain  to scale
$l$ and replace $w$ by $w(l)$,  $r$ by $r(l)$ and $h^2$ by $h(l)^2$ to
obtain   $h(l)^2  \sim   w(l)\,|r(l)|^{d/2-1}$. Eq. (\ref{largel}) shows 
that $w(l)$ becomes progressively smaller as $l$ increases.  
Our use of the perturbative result will  become valid at some value  
$l=l^*$, where $l^*$ will be specified below. 
Inserting the  $l$ dependent  forms for $h(l)$,  $w(l)$ and  $r(l)$ at
$l=l^*$, one obtains
\begin{equation}
h(0)^2\equiv
h^2_{AT}=\frac{w(0)|r(0)|^{d/2-1}}{\left[(2w(0)^2/\epsilon)(1-e^{-\epsilon
l^*})+1\right]^{5d/6-1}}.
\label{match}
\end{equation}
 Setting  $d=6+\epsilon$,  and  taking  $\exp(-\epsilon  l^*)\ll1$  we
obtain the final result, correct to leading order in $\epsilon$, that
\begin{equation}
h^2_{AT}\sim                    \left(\frac{\epsilon}{2w(0)^2}\right)^4
w(0)|r(0)|^{d/2-1}     \equiv     \left(\frac{\epsilon}{2w^2}\right)^4
w|r|^{d/2-1}.
\label{srfinal}
\end{equation}
This is equivalent to Eq.  (\ref{final result}).

One cannot  simply set  $l^*\rightarrow \infty$ in  Eq. (\ref{match}).
The RG equations as presented here are valid provided $r(l^*)$ remains
small or comparable to the square of  the cut-off in the  theory, which is
conventionally    taken    to    be    unity.   To    leading    order
$r(l^*)=r(0)\exp(2l^*)$,  so we shall fix  $l^*$ by setting 
\cite{Fisher-Sompolinsky}
\begin{equation}
|r(l^*)| \approx |r(0)|e^{2l^*}=1.
\label{l*1}
\end{equation}
 Eq. (\ref{srfinal}) will  hold provided that
\begin{equation}
e^{-\epsilon l^*}=|r(0)|^{\epsilon/2}\equiv|r|^{\epsilon/2}\ll 1.
\label{limit}
\end{equation}
Thus as $\epsilon \rightarrow  0$, the temperature interval near $T_c$
over which Eq. (\ref{final result}) is accurate becomes very narrow.

\section{Does RSB disappear for $d<6$?}

We  are arguing in this paper that 
the  lower critical dimension for the  AT line -- the
line  beneath which  RSB sets  in in  the presence  of a  field  -- is
$6$. In zero field there is  still a phase transition when $d\le 6$,   
provided $d>d_l$, where $d_l$ is the lower critical dimension, thought 
to be between 2 and 3. One might  wonder whether  replica  symmetry 
breaking  is present in  the zero-field phase  or whether RSB too 
disappears  below six dimensions.
We  next  present an  argument  that  RSB  in zero field  vanishes  as
$d\rightarrow 6^+$ based  on the same methods which  we used to derive
the dimensionality dependence of the AT line.

When $d>6$ the form of replica symmetry breaking is well-established
\cite{Parisi, DKT}.  $\langle Q_{\alpha  \beta}\rangle$ becomes  a function
$q(x)$ in the  interval $1\ge x\ge 0$ , which is  constant for $1\ge x
\ge x_1$  at the value  $q_{EA}\,\,(=Q)$ and then falls  from
this value to zero at $x=0$.  For the functional of Eq. (\ref{FQ}) (with $h^2$ set to zero)  the breakpoint $x_1$ is given to one-loop order for $6<d<8$ by
\cite{DKT}
\begin{equation}
x_1 \sim  w^2|r|^{d/2-3}\,.
\label{bp}
\end{equation}
If $x_1$  were zero, the spin  glass would be
replica symmetric.  We shall argue  that as $d \rightarrow 6^+$, $x_1$
goes to zero  linearly with $(d-6)$, suggesting that  when $d \le 6$ 
there will be no replica symmetry breaking.
 Higher loop terms will leave the
exponent of $|r|$ unchanged, but modify its prefactor, just as they do
for the AT line of Eq. (\ref{AT1}).

We shall work in  $6+\epsilon$ dimensions again with $\epsilon$ small
and take $|r|$ small  so as to permit neglect of the quartic terms in the 
functional 
 Eq.  (\ref{FQ}). The RG will be used to coarse-grain to scale
$l^*$ at which the form  of the perturbative one-loop order expression
for the breakpoint $x_1$ becomes valid.  The breakpoint $x_1[r,w]$ has
zero scaling dimension, so it follows that
\begin{equation}
x_1[r,w]=x_1[r(l^*),w(l^*)] \sim w(l^*)^2 |r(l^*)|^{d/2-3}.
\end{equation}
Inserting  our previous  expressions  for $r(l^*)$  and $w(l^*)$,  one
finds
\begin{equation}
x_1[r,w]\sim             \frac{w(0)^2|r(0)|^{d/2-3}}{[(2w(0)^2/\epsilon
)(1-e^{-\epsilon l^*})+1]^{5d/6-4}}.
\end{equation}
 $l^*$ is  specified as
in Eq. (\ref{limit}).
Hence in the limit when $\epsilon$ is small (but $\epsilon l^*\gg 1$)
\begin{equation}
x_1\sim \frac{1}{2} (d-6)|r|^{d/2-3},
\end{equation}
which goes  to zero as $d  \rightarrow 6^+$, implying  that replica
symmetry breaking vanishes in  this limit. 

\section{The Bray-Roberts calculation and the Island of Stability}

We shall now describe the BR calculation of the RG equations pertinent
to the AT  line. It is this work  which is at the heart  of our contention
that $6$ is  the lower critical dimension for the  existence of the AT
line.  As mentioned before, their  approach is to study just the fields
in the replicon sector $\tilde{Q}_{\alpha \beta}$, which are such that
$\sum_{\beta}\tilde{Q}_{\alpha \beta}=0$. The effective functional is
\begin{eqnarray}
&F[\{\tilde{Q}_{\alpha\beta}\}]  =  \int   d^dx\,   \left[\frac{1}{4}\tilde{r}
\sum \tilde{Q}_{\alpha\beta}^2
+\frac{1}{4}\sum(\nabla
\tilde{Q}_{\alpha\beta})^2 \right. \nonumber   \\ &   + \left.(w_1/6)
\sum \tilde{Q}_{\alpha\beta}
\tilde{Q}_{\beta\gamma}\tilde{Q}_{\gamma\alpha} 
+(w_2/6)\sum \tilde{Q}_{\alpha\beta}^3\right]. 
\label{FBR}
\end{eqnarray}
Here the convention has been adopted that the sums over replica indices 
are unrestricted except that $\tilde{Q}_{\alpha \alpha}=0$. At the AT line,
$\tilde{r} =0$ in the mean-field approximation. According to BR, the 
coupling constants
$w_1$ and $w_2$ would have  for $d>6$ the  RG flow equations 
\begin{equation}
\frac{dw_1}{dl}  =  \frac{1}{2}
\left[-\epsilon -3\eta_{R}(l)\right]w_1+14w_1^3-36w_1^2w_2+18w_1w_2^2+w_2^3
\label{ATRG1}
\end{equation}
\begin{equation}
\frac{dw_2}{dl}  =  \frac{1}{2}\left[-\epsilon -3\eta_{R}(l)\right]
w_2+24w_1^2w_2-60w_1w_2^2+34w_2^3.
\label{ATRG2}
\end{equation}
Once again we  have adopted the convention of  absorbing the geometric
factor    $K_d$    into   $w_1$    and    $w_2$   and    $\eta_{R}(l)=
(4w_1^2-16w_1w_2+11w_2^2)/3$.   Presumably   one  could  obtain  these
replicon  sector  equations  by  integrating  the  full  equations  of
Ref. \cite{Temesvari}  containing the hard  longitudinal and anomalous
modes as well as the replicon modes until the hard modes are decoupled
from those in the replicon sector.  If one were able to carry out this
formidable  task  the initial  values  of  $w_1$  and $w_2$  could  be
determined.  We suspect that the initial value of $w_1$ would turn out
to  be of  order $w(l^*)  ( \sim  \epsilon^{1/2}|r|^{\epsilon/4})$. At
mean-field level  $w_2$ is $\sim yQ$,  where $y$ is
the     coefficient     of      a     particular     quartic     term,
$(y/4)\sum_{\alpha<\beta}Q^4_{\alpha \beta}$, found when one goes beyond the
cubic functional  in Eq.   (\ref{FQ}) \cite{Bray-Roberts}. The effective 
value of $y$ is dominated at small $|r|$ when $6<d<8$ by its  renormalization  
by the four  cubic
vertex \lq\lq  box'' diagram \cite{Fisher-Sompolinsky}. Then  $y \sim w^4
|r|^{d/2-4}$ and so     $w_2\sim
w(l^*)^3|r(l^*)|^{\epsilon/2}  (\sim \epsilon^{3/2}|r|^{3\epsilon/4})$.
Fortunately we  do not  need the precise  initial values of  $w_1$ and
$w_2$ for our  argument, which is based upon the form  of the basin of
attraction      of      the      Gaussian     fixed      point      of
Eqs. (\ref{ATRG1})-(\ref{ATRG2}).

\begin{figure}
\includegraphics[height=5.0cm,width=7.5cm]{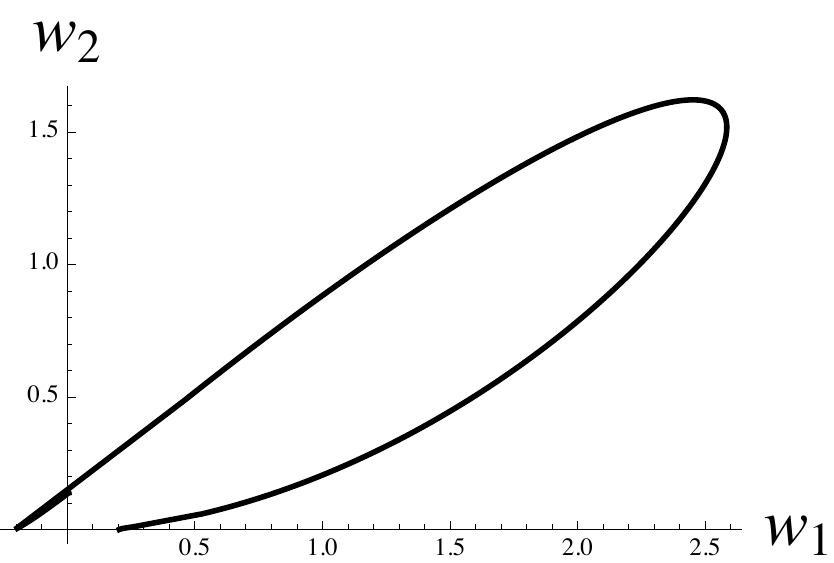}
\caption{\label{fig:1} The island of stability of the Gaussian fixed 
point for $d>6$. Distances are measured in units of $\sqrt{\epsilon}$.
Only the region $w_2>0$ is displayed.} 
\end{figure}

The basin of attraction has been determined numerically, and is displayed 
in Figure 1. By scaling both $w_1$ and $w_2$ by $\epsilon^{1/2}$  and 
writing $\tilde{l}=\epsilon  l$, the explicit dependence on $\epsilon$ in  
Eqs. ({\ref{ATRG1})-(\ref{ATRG2}) can be removed. There is a  peculiarly 
shaped compact region (the ``island of stability'')  around the origin  
in the  $w_1,w_2$  space inside  which  all flows  are to  the Gaussian 
fixed  point $w_1^*=0=w_2^*$.  Outside this region  the flows
are to  infinity.  The  linear extent of  this basin of  attraction is
from the scaling of $w_1$ and $w_2$ of order $\epsilon^{1/2}$. Thus as
$\epsilon \rightarrow 0$, the size  of the basin of attraction shrinks
to  zero.  Now  BR showed  that when  $d \le  6$ there  was  no stable
physical  fixed  point;   all  flows  of  $w_1$  and   $w_2$  were  to
infinity. Since the basin of  attraction of the fixed point associated
with the  AT line  is shrinking  to  zero as  $d \rightarrow  6^+$ and  
no physical stable fixed points exists for $d \le 6$, it seems natural to
expect that  the lower critical dimension  of the AT  line and replica
symmetry breaking generally must be $6$. This is a result in line with 
earlier expectations \cite{Bray-Moore} but is also in accord with more 
recent results for the \lq\lq strongly-disordered spin-glass model", 
a model with an  unusual  bond-distribution  \cite{JR}.

 The behavior of  the basin of attraction of  the Gaussian fixed point
 of Eqs.\,(\ref{ATRG1})-(\ref{ATRG2}) as $d \rightarrow 6^+$ should
 be contrasted with that of  Eq.\,(\ref{w-eq}) for the zero-field case.
 In  the latter  case  the basin  does not shrink to zero as 
$d \rightarrow 6^+$ and there is a phase transition  when $d_l <d \le 6$.

 Our  treatment  of  the AT  line  was  done  using the  zero-field  RG
 equations with one coupling constant  $w$ rather than the full set of
 coupling       constants,        $w_1,w_2,       \cdots,w_8$       of
 Ref. \cite{Temesvari}. The justification for this is that  all but 
$w_1$  can be dropped
when obtaining the form of the AT line as $T\rightarrow T_c$. To elaborate 
this point further we have already quoted  forms for $w_2(l^*)$ and $w_1(l^*)$ 
and as  $|r|\rightarrow 0$,  $w_2(l^*)$ is indeed neglible in comparison 
with $w_1(l^*)$. However  for $d\le 6$, the  situation is
 quite different. If the coupling coefficients $w_2, w_3, \cdots, w_8$
 are anything but zero, the RG flows will take them to infinity and so
for $d\le 6$ 
 it is never  possible to work with just  the one-coupling constant RG
 equations.  The same point is relevant  for the   attempts to extend 
RSB calculations to $d \le 6$ in zero-field \cite{DKT}.   For $d>6$  
all the  coupling constants  can flow  to the Gaussian fixed point.
 
\section{The One-Dimensional Spin Glass with Long-Ranged Interactions}

Because our calculation on the form of the AT line is only valid when
$\epsilon$  is small,  there  is no  chance  that it  can be  directly
checked by simulations.  However, we can derive the analogous results 
for the one-dimensional Ising spin glass with long-range interactions, 
whose AT line has recently been the subject of contradictory numerical 
simulations \cite{Larson,Leuzzi2009}. The Hamiltonian of these studies 
are variants of 
\begin{equation}
H=-\sum_{<ij>}J_{ij}S_iS_j -\sum_i h_iS_i,
\label{Ham}
\end{equation}
where $h_i$ is a random field of variance $h^2$, the  sum is
over  all  pairs  $<ij>$,  and  $i$  and $j$  are  positions  on  the
 one-dimensional lattice. The interaction
\begin{equation}
J_{ij}=J\frac{\epsilon_{ij}}{|i-j|^{\sigma}},
\label{sigma}
\end{equation}
where  the $\epsilon_{ij}$  are  independent random  variables with  a
Gaussian distribution of  zero mean and unit variance.  This model was
introduced by Kotliar et al.  \cite{KAS}, who showed that for $\sigma<
2/3$ the  model has mean-field critical exponents,  and non-mean field
exponents  for $2/3  <\sigma <  1$.  When  $\sigma >  1$, there  is no
finite  temperature  phase  transition.  The  $\sigma$-interval,  $2/3
<\sigma  < 1$, is  the analogue  for short-range  spin glasses  of the
dimension range between the upper critical dimension ($d_u=6$) and the
lower  critical   dimension,  while  $\sigma   <2/3$  corresponds  to
dimensionalities $d>6$.  Thus just by  changing the value  of $\sigma$
one  can   explore  both  systems   corresponding  to  high   and  low
dimensionality.

Our expectation is that when $\sigma<2/3$ there will be an AT line and
both Refs.\ \cite{Larson,Leuzzi2009} confirm this.  Unfortunately for
$\sigma >2/3$ the  two groups of simulators were  in disagreement with
each other: Ref.\ \cite{Larson} did not see an  AT transition, whereas
Ref.  \cite{Leuzzi2009} did.  Our prediction below of the  form of the
AT  line as  $\sigma \rightarrow  2/3^-$ supports  the  conclusions of
Ref. \cite{Larson}.

The RG  equations near  the upper critical  value of  $\sigma$, $2/3$,
were first  written down in  Ref. \cite{KAS}. They exploited  the fact
that for these long-range interactions $2-\eta=2\sigma-1\,$\cite{Sak}.
The bare  propagator is  $1/(q^{2\sigma-1}+r)$. The RG  flow equations
become
\begin{eqnarray}
\frac{dw}{dl} & = & -(2-3\sigma)w-2w^3 
\label{lrw-eq} \\
\frac{dr}{dl} & = & (2\sigma-1)r - 4w^2 r,
\label{lrr-eq}
\end{eqnarray}
while $h^2$  grows as
\begin{equation}
\frac{d(h^2)}{dl} = \sigma h^2.
\label{lrh-eq}
\end{equation}
Eqs. (\ref{lrw-eq}) and (\ref{lrr-eq}) are valid only when $w$ is small, 
but Eq. (\ref{lrh-eq}) is exact.

The perturbative calculation along the lines of Green et al. 
\cite{Green-Moore-Bray} of the AT line to one loop order gives a result 
similar to  Eq. (\ref{GBMFS}):
\begin{equation}
h^2/Q = 144 w^2 |r|^2 I_{\sigma}\ , 
\label{jenny2}
\end{equation}
where $I_{\sigma}=\int_0^{\infty} dq\, q^{-2(2\sigma-1)}(q^{2\sigma-1}
 + |r|)^{-2}$, which equals $B_{\sigma}|r|^{(5-8\sigma)/(2\sigma-1)}$, 
and $B_{2/3}=3$. Hence to one-loop order the equation of the AT line is
\begin{equation}
h^2 \sim w|r|^{\frac{2-2\sigma}{2\sigma-1}},\ \ \ (2/3 > \sigma >5/8).
\label{lrGBMFS}
\end{equation} 
For $\sigma \rightarrow 5/8$, the mean-field AT form 
$h^2 \propto |r|^3$ is recovered.  Hence $\sigma =5/8$ is the analogue 
of 8 dimensions for short-range spin glasses.

It is straightforward to integrate the RG equations for the long-range 
case  to obtain the form of the 
AT line as $\sigma \rightarrow 2/3^-$.  The result is 
\begin{equation}
h(0)^2 \sim
\left(\frac{2-3\sigma}{w(0)^2}\right)^{\frac{3-2\sigma}{2(2\sigma-1)}} 
w(0)|r(0)|^{\frac{2-2\sigma}{2\sigma-1}} 
\end{equation}
or equivalently
\begin{equation}
\frac{h^2_{AT}}{T_c^2}=C(2-3\sigma)^{\frac{3-2\sigma}{2(2\sigma-1)}} 
\left(1-\frac{T}{T_c}\right)^{\frac{2-2\sigma}{2\sigma-1}},
\label{lrfinal}
\end{equation}
and  $C$ is again a constant of $O(1)$. Hence  as $\sigma\rightarrow 2/3^-$ 
the AT line goes away. That there was no AT transition in the interval 
$1>\sigma>2/3$ has also recently been suggested \cite{Moore2010} from an 
expansion about $\sigma=1$, the \lq\lq lower critical value".

The equivalent result for the ``break-point'', $x_1$, in the Parisi function 
for the one-dimensional long-ranged system is 
\begin{equation}
x_1 \ \sim (2-3\sigma) |r|^{(4-6\sigma)/(2\sigma-1)}, 
\end{equation}
when $\sigma \to 2/3-$.

\section{Conclusion}

We have presented arguments that the Almeida-Thouless line in 
spin glasses is absent in systems with six or fewer space 
dimensions, i.e.\ these systems exhibit no phase transition under 
cooling if an external magnetic field is present. We have also argued 
that the features associated with broken replica symmetry, in the Parisi 
solution of the SK model, are not present in finite-dimensional spin 
glasses with $d \le 6$.  

Equivalent results have been obtained for a one-dimensional spin glass 
with interactions decaying with distance $r$ as $1/r^\sigma$. These systems 
are more amenable to simulation than short-ranged systems in high-dimensional 
space. To  date, however,  there  seem   to  be  no   simulational  
studies   of  the one-dimensional   long-range  model  which   might  help   
to  confirm Eq. (\ref{lrfinal}) in the  interval  $5/8<\sigma<2/3$. 
One issue which will complicate  such studies was  pointed out in  
Ref. \cite{Larson}.
Simulations at  the AT line require  the system to be  large enough so
that   the  Parisi   overlap   function  $P(q)$   has  only   positive
support. This requires  $h (QN)^{1/2}>T$. Whether simulations can
be  done with  the  number of  spins  $N$ large  enough  to meet  this
requirement remains to be seen.

The  debate  as  to the  nature  of  the  spin  glass phase  in  three
dimensional systems  has run for  so long because on  the experimental
side, for example,  dynamical effects can produce an  apparent AT line
\cite{Bray-Moore,Mattson}}, while  in simulations  there  are always
finite size  effects which  can mimic some  of the effects  of replica
symmetry breaking \cite{FSBM}.  As a  consequence it 
was always hard to
be  certain which of  the two  pictures, RSB  or droplet  scaling, was
correct.  Ref.\,\cite{JKK} is an example of a 
recent simulation and provides further references.  We  believe that  
our calculations provide  strong arguments that the spin  glass phase 
will not have  replica symmetry breaking in dimensions $d \le 6$. 

\section{Acknowledgements}

We are indebted to G. Parisi, T. Temesvari, and A. P. Young  for 
their useful comments.

\end{document}